# Coherent soft-mode phonon generation and detection in ultrathin $SrTiO_3$ grown directly on silicon

Cheng Cen[1], Maitri P. Warusawithana[2], Darrell G. Schlom[3], Jeremy Levy[1]

[1]Department of Physics and Astronomy, University of Pittsburgh, Pittsburgh, PA 15260

[2]Department of Physics, Florida State University, Tallahassee, FL 32310

[3]Department of Material Science and Engineering, Cornell University, Ithaca, NY 14853

**Abstract**

Time-resolved two color pump-probe polarization spectroscopy was performed at room temperature on $SrTiO_3$ films grown directly on Si with film thickness varying from 2 nm to 7.8 nm. The *E* soft mode with a characteristic frequency of 0.2 THz is impulsively generated and measured in these coherently strained tetragonal phase $SrTiO_3$ thin films. Another over-damped signal observed indicates the possible relaxational hopping of Ti ion between double potential wells. The dependence of the coherent phonon signal on pump and probe laser polarization helps to identify the phonon modes.

Great effort has been made in the integration of traditional semiconductors with oxide materials that exhibit numerous novel functional properties such as ferroelectricity [1, 2], superconductivity [3] and a field-tunable interfacial metal insulator transition [4, 5]. Among all oxide/semiconductor interfaces, high quality $SrTiO_3$/Si heterostructures are especially intriguing, because $SrTiO_3$ combines most of the functionality of oxide materials [1-5], and can serve as a template for the growth of various other oxide thin films [6-9]. Successful growth of strain-engineered ferroelectric $SrTiO_3$ thin films directly on Si was reported recently [2]. For future material quality improvement and ferroelectric device implementations, it is critical to understand the lattice dynamics in the epitaxial oxide. Of particular importance is the lowest-frequency transverse optical phonon, also known as the soft mode, whose frequency is related to the static dielectric constant by the Lyddane-Sachs-Teller (LST) relationship and approaches zero at the paraelectric-to-ferroelectric transition temperature [10].

Traditionally, frequency domain methods such as Raman scattering and neutron diffraction have been used to detect phonon oscillations [11-17]. Measurement of soft modes in bulk $SrTiO_3$ is very challenging due to the Raman inactive nature and the relatively low frequencies. Epitaxial growth of $SrTiO_3$ ultrathin films is capable of introducing lattice strain which can help the observation of soft modes by breaking the central symmetry of the lattice and inducing soft-mode hardening [13, 14, 18, 19]. However, the extremely small material thickness also suppresses the scattering cross section. On the other hand, time domain pump-probe techniques, which use femtosecond laser sources to generate and detect spatially and temporally coherent lattice vibrations, have been proven quite powerful for the investigation of both semiconductors and complex oxide systems [20-22].

Here, we present room-temperature measurements of a soft optical phonon mode in strained $SrTiO_3$ films grown directly on Si. We employ a two-color pump-probe electro-optic sampling method to impulsively stimulate and detect coherent phonons in ultrathin films of $SrTiO_3$ grown directly on Si. The electrooptic contrast arises due to a transient anisotropy in the material's refractive index caused by the excitation of the coherent phonons. Because of the high signal-to-noise ratio of the two-color balanced detection technique, we are able to observe clear soft phonon oscillations in films as thin as $d = 5$ ML (2 nm), where ML is the number of monolayers of (001)-oriented $SrTiO_3$, each a unit cell thick, comprising the film.

The $SrTiO_3$ films of thickness $d = 5, 6, 8, 10, 20$ ML studied here were grown by molecular-beam epitaxy (MBE) with epitaxial orientation: (001)$SrTiO_3$ // (001)Si; [110]$SrTiO_3$ // [100]Si. A detailed description of the growth method is described elsewhere [2]. This orientation relationship gives rise to a uniform biaxial compressive strain in $SrTiO_3$, which is proved to be critical in inducing the ferroelectricity [2].

Coherent phonon dynamics at room temperature are investigated with ultrafast polarization spectroscopy using a two-color pump-probe method (Fig. 1(a)). A linearly polarized pump pulse (120 fs) is intensity modulated at $\omega/2\pi = 42$ kHz and then focused at normal incidence onto the samples. The pump pulse (center wavelength $\lambda_{pump} = 820$ nm) imparts an impulsive force to the sample lattice, creating coherent phonons through an impulsive stimulated Raman scattering (ISRS) process [23, 24] . These coherent phonons are probed using a probe pulse generated at the second harmonic of the pump beam (center wavelength $\lambda_{probe}$ = 410 nm). This probe beam is time-delayed and focused to the same ~1 μm spot on the sample. Phonon oscillations induce periodic anisotropy in the refractive index of the sample and thus change the polarization of the reflected probe beam. The delay time between pump and probe is swept to map out the time-resolved signal. The reflected probe beam is split into signal and reference channels of a balance detector using a polarizing beam splitter. To suppress the unwanted background and noises, the balance is adjusted so that the signal before zero delay is minimized. To exclude the influence of mechanical vibrations caused by the photoelastic modulator, the polarization rotation of probe pulse is measured at the second harmonic of the pump's modulation frequency.

With the incident probe beam's polarization perpendicular to the pump, signals of polarization change $\Delta I$ of the reflected probe beam are measured as a function of the delay time $t$ between the pump and probe pulses (Fig. 1(b)). Clear coherent oscillations can be seen preceded by a sharp increase of signal amplitude at zero delay. To make sure this signal is solely from the ultrathin $SrTiO_3$ films, a control experiment using the identical experimental setup is performed on a bare Si substrate and no comparable time-dependent signal was observed. The time dependent signal $\Delta I$ is not affected by the change in the polarization of the pump beam relative to the sample, but is strongly dependent on the angle $\theta$ between the polarizations of the incident pump and probe pulses. At $\theta = 45^\circ$, the phonon oscillation signal completely vanishes; at $\theta = 0^\circ$ and $90^\circ$, the amplitude is maximized, while the phase differs by $\pi$ (Fig. 2). The total signal $\Delta I$ can be treated as the sum of a vibrational mode $\Delta I_1$ and an over-damped mode $\Delta I_2$, which decays away in a few picoseconds, both of which will discussed below.

The vibration mode $\Delta I_1$ measured in films with different thicknesses are presented in Fig. 3(a). The overall signal profile is consistent with a damped harmonic oscillation model, $\Delta I_1 = A_1 e^{-t/\tau_1} \sin(2\pi f_1 t)$. A single frequency peak at $f_1$ = 0.2 THz can be seen in the Fourier transform of $\Delta I_1(t)$ (Fig. 3(b)). This frequency is much lower than the reported hard optical phonon frequencies in unstrained $SrTiO_3$ [25-27]. One possible low frequency excitation is an acoustic phonon. However, zone center acoustic modes created in first order Raman processes have much lower frequencies. Acoustic modes with large wave vectors can be generated in second order processes, but the phonon frequencies are expected to be film thickness dependent. This contradicts the observation that $f_1$ remains constant for all the film thicknesses measured. In addition, the film thicknesses are much less than the acoustic wavelength $\lambda_a = \lambda_{probe} / 2n_{STO} > 89\text{nm}$ (the refractive index of $SrTiO_3$ thin film $n_{STO} < 2.3$ [28]) required by the phase-matching condition. The dependence of the phonon oscillations on the polarization angle between pump and probe further rules out the possibility of having acoustic waves traveling in-plane generated by electron heating caused by an ultrashort pump pulse [29]. On the other hand, at room temperature, $SrTiO_3$ films biaxially strained by commensurate growth on a Si substrate are in a tetragonal phase [2] in which the soft mode splits into phonons with $E$ and $A$ symmetries. Although Raman inactive in bulk, soft optical modes have been observed in $SrTiO_3$ with frequencies comparable to $f_1$ under the presence of strain or electric field [18, 30]. Therefore, we attribute the 0.2 THz phonon oscillations to a low frequency *E* soft optical mode, which is made Raman active by the epitaxial strain present in the films.

We note that the polarization angle dependence discussed previously is consistent with the symmetry of *E* soft mode. When the probe polarization has a 45 degree angle with the pump, the two degenerate *E* phonon modes rotate the polarization of the scattered probe pulse by the same amount in opposite directions, giving rise to a zero net modulation of the probe pulse's polarization.

In thicker films, larger oscillation amplitudes are measured because of a larger scattering cross section between light and lattice (Fig. 3(c)). The extraction of decay constant $\tau_1$ yields a mean value of 27.7ps and a less than 4% variation in the films with different thicknesses ($\tau_1$ is not calculated for the 5 ML film due to the relatively low signal-to-noise ratio) (Fig. 3(c)). Signatures of small deviations from a single damped sinusoidal function are observed (Fig. 3(a), arrow pointed). These features are repeatable in experiments performed at different spots on the films and should not be confused with common noise. The possible existence of slightly relaxed portion of $SrTiO_3$ film, as was seen by high resolution x-ray diffraction measurements [2], may contribute phonon oscillations with a different frequency and give rise to these beating-like waveforms.

Figure 4(a) shows the other over-damped signal component $\Delta I_2$. The several picosecond long life time of $\Delta I_2$ is more than one order of magnitude longer than the pulse width of the pumping and probing laser, therefore it cannot be attributed to a commonly observed coherent artifact [31]. It is suggested that the hopping of ions between the double potential minima in ferroelectric perovskites can lead to an over-damped relaxation mode [32]. Relaxation modes have been observed in several materials by Raman spectroscopy [33, 34] and ISRS experiments [20, 35] with typical decay constants ranging from a few picoseconds to nanoseconds. Although bulk $SrTiO_3$ is paraelectric down to zero temperature, the strained films we measured with thickness $d$=5,6,8,10 ML have been verified to be ferroelectric at room temperature by piezoforce microscopy [2]. The measured over-damping signal $\Delta I_2$ may reflect such a relaxation mode in ferroelectric $SrTiO_3$ films. The decreasing decay constant $\tau_2$ in thicker films (Fig. 4(b)) corresponds to a higher hopping rate compared to the dwell time of Ti ion in each of the potential wells. This effect represents the possible reduction of the confinement potential that holds the bistable Ti ion in each individual site as more $SrTiO_3$ layers are grown. Indeed, no successful writing or imaging of ferroelectric domains was achieved by piezoforce microscopy in the film with $d$=20 ML [2].

In summary, we have observed coherent phonon modes in ultra-thin $SrTiO_3$ films grown directly on Si substrates using an ultrafast pump-probe method. The underdamped vibrational modes are believed to belong to the $E$ symmetry soft phonon, while the other over-damped signal is attributed to a relaxation mode related to the bistable ion potential profile in ferroelectric films. The phonon dynamics probed directly by ISRS measurements provide useful information about the ferroelectric properties in these Si strained $SrTiO_3$ films.

We acknowledge helpful discussions with H. Petek. This work is supported by NSF-DMR-0704022 (JL) and AFOSR through Award No. FA9550-10-1-0524 (DGS).

[1] J. H. Haeni *et al.*, Nature **430**, 758 (2004).
[2] M. P. Warusawithana *et al.*, Science **324**, 367 (2009).

[3] J. F. Schooley, W. R. Hosler, and M. L. Cohen, Physical Review Letters **12**, 474 (1964).
[4] C. Cen *et al.*, Nature Materials **7**, 298 (2008).
[5] C. Cen *et al.*, Science **323**, 1026 (2009).
[6] P. Chaudhari, F. K. Legoues, and A. Segmuller, Science **238**, 342 (1987).
[7] N. D. Mathur *et al.*, Nature **387**, 266 (1997).
[8] N. Reyren *et al.*, Science, 1146006 (2007).
[9] R. Ramesh *et al.*, Science **252**, 944 (1991).
[10] W. Cochran, Adv. Phy. **9**, 387 (1960).
[11] P. A. Fleury, J. F. Scott, and J. M. Worlock, Physical Review Letters **21**, 16 (1968).
[12] A. A. Sirenko *et al.*, Physical Review Letters **82**, 4500 (1999).
[13] A. A. Sirenko *et al.*, Nature **404**, 373 (2000).
[14] I. A. Akimov *et al.*, Physical Review Letters **84**, 4625 (2000).
[15] D. A. Tenne *et al.*, Physical Review B **70** (2004).
[16] S. M. Shapiro *et al.*, Physical Review B **6**, 4332 (1972).
[17] K. Hirota *et al.*, Physical Review B **52**, 13195 (1995).
[18] H. Uwe, and T. Sakudo, Physical Review B **13**, 271 (1976).
[19] I. Fedorov *et al.*, Ferroelectrics **208**, 413 (1998).
[20] T. P. Dougherty *et al.*, Science **258**, 770 (1992).
[21] G. A. Garrett *et al.*, Science **275**, 1638 (1997).
[22] K. J. Yee *et al.*, Physical Review Letters **88** (2002).
[23] Y. R. Shen, and N. Bloembergen, Phys. Rev **137**, 1787 (1965).
[24] Y. X. Yan, and K. A. Nelson, The Journal of Chemical Physics **87**, 6240 (1987).
[25] W. G. Nilsen, and J. G. Skinner, The Journal of Chemical Physics **48**, 2240 (1968).
[26] H. Vogt, Physical Review B **38**, 5699 (1988).
[27] H. Vogt, and G. Rossbroich, Physical Review B **24**, 3086 (1981).
[28] M. Wohlecke, V. Marrello, and A. Onton, Journal of Applied Physics **48**, 1748 (1977).
[29] I. Bozovic *et al.*, Physical Review B **69** (2004).
[30] J. M. Worlock, and P. A. Fleury, Physical Review Letters **19**, 1176 (1967).
[31] A. J. Taylor, D. J. Erskine, and C. L. Tang, J. Opt. Soc. Am. B **2**, 663 (1985).
[32] M. E. Lines, and A. M. Glass, *Principles and Applications of Ferroelectrics and Related Materials* (Clarendon, Oxford, 1977).
[33] M. D. Fontana *et al.*, J. Phys. C: Solid State Phys **21**, 5853 (1988).
[34] J. P. Sokoloff, L. L. Chase, and D. Rytz, Physical Review B **38**, 597 (1988).
[35] T. P. Dougherty *et al.*, Physical Review B **50**, 8996 (1994).

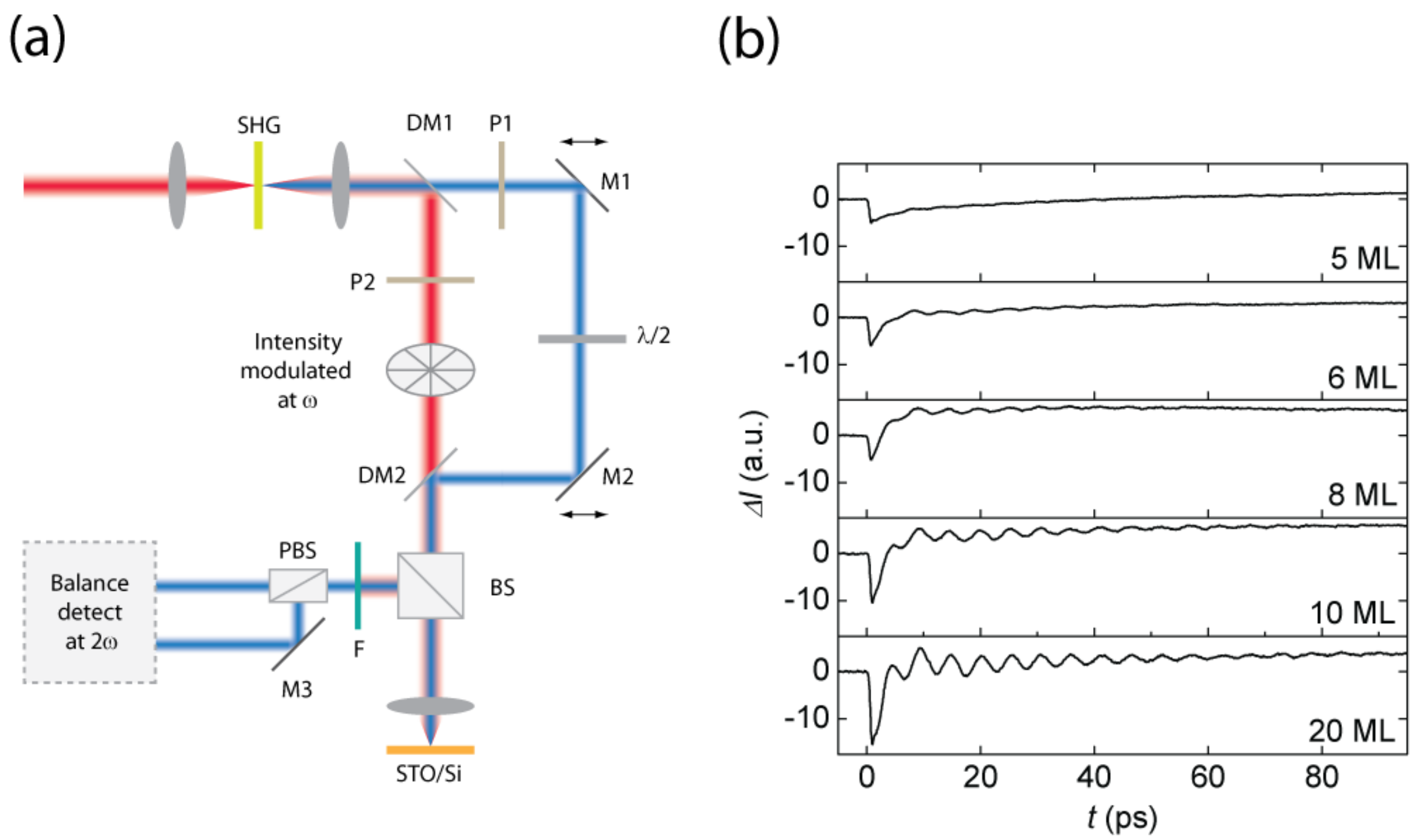

**Figure 1** (a) Layout of pump-probe experiment setup. SHG: second harmonic generator; DM: dichroic mirror; P: polarizer; M: mirror; λ/2: half wave plate; BS: beam splitter; F: filter; PBS: polarizing beam splitter. (b) Coherent phonon oscillations measured in $SrTiO_3$/Si films (STO/Si) with different thicknesses.

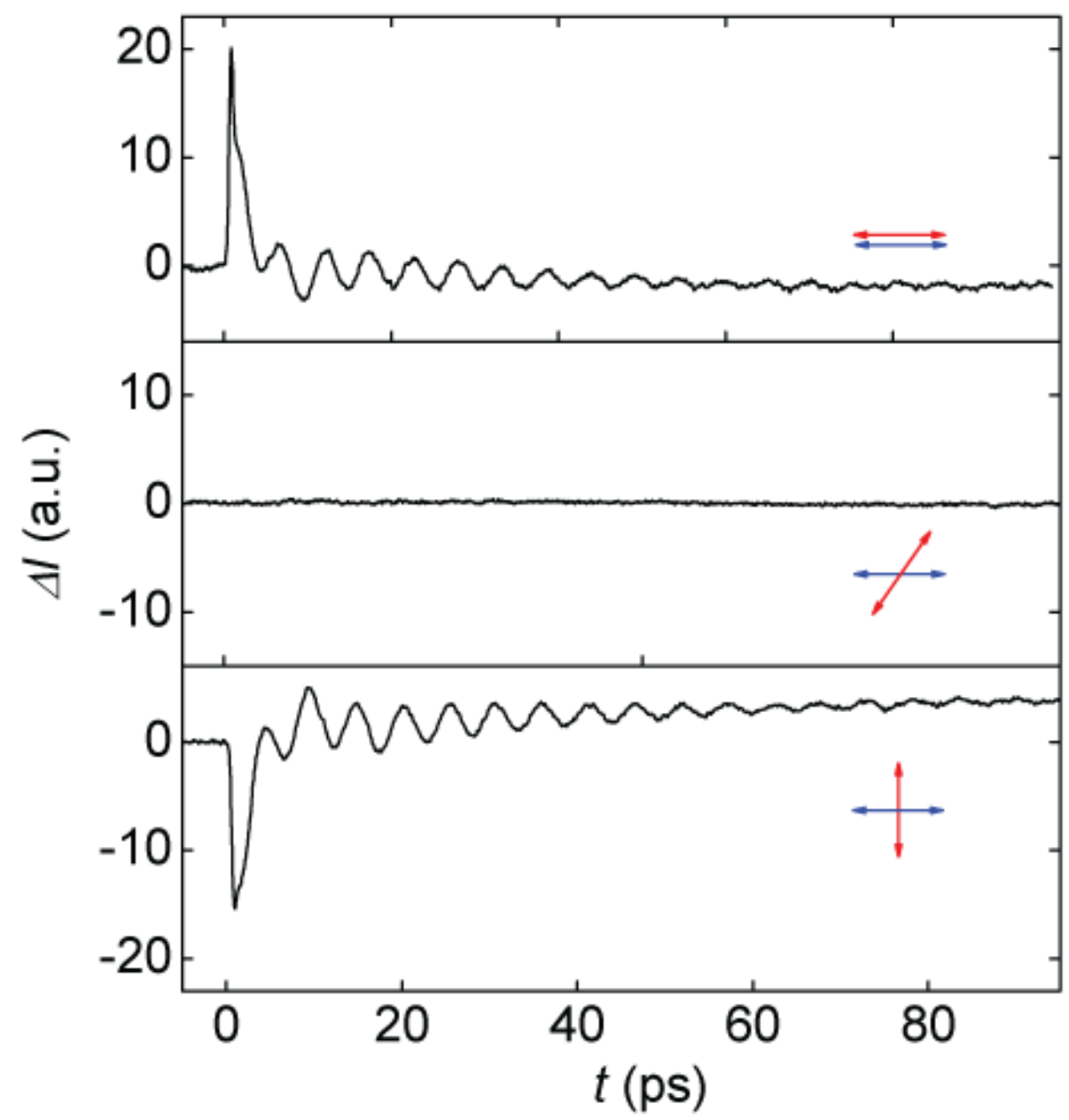

**Figure 2** Phonon oscillations in a 20 ML thick $SrTiO_3$/Si film at three different relative polarization angles between pump and probe $\theta = 0^{\circ}$, $45^{\circ}$, $90^{\circ}$.

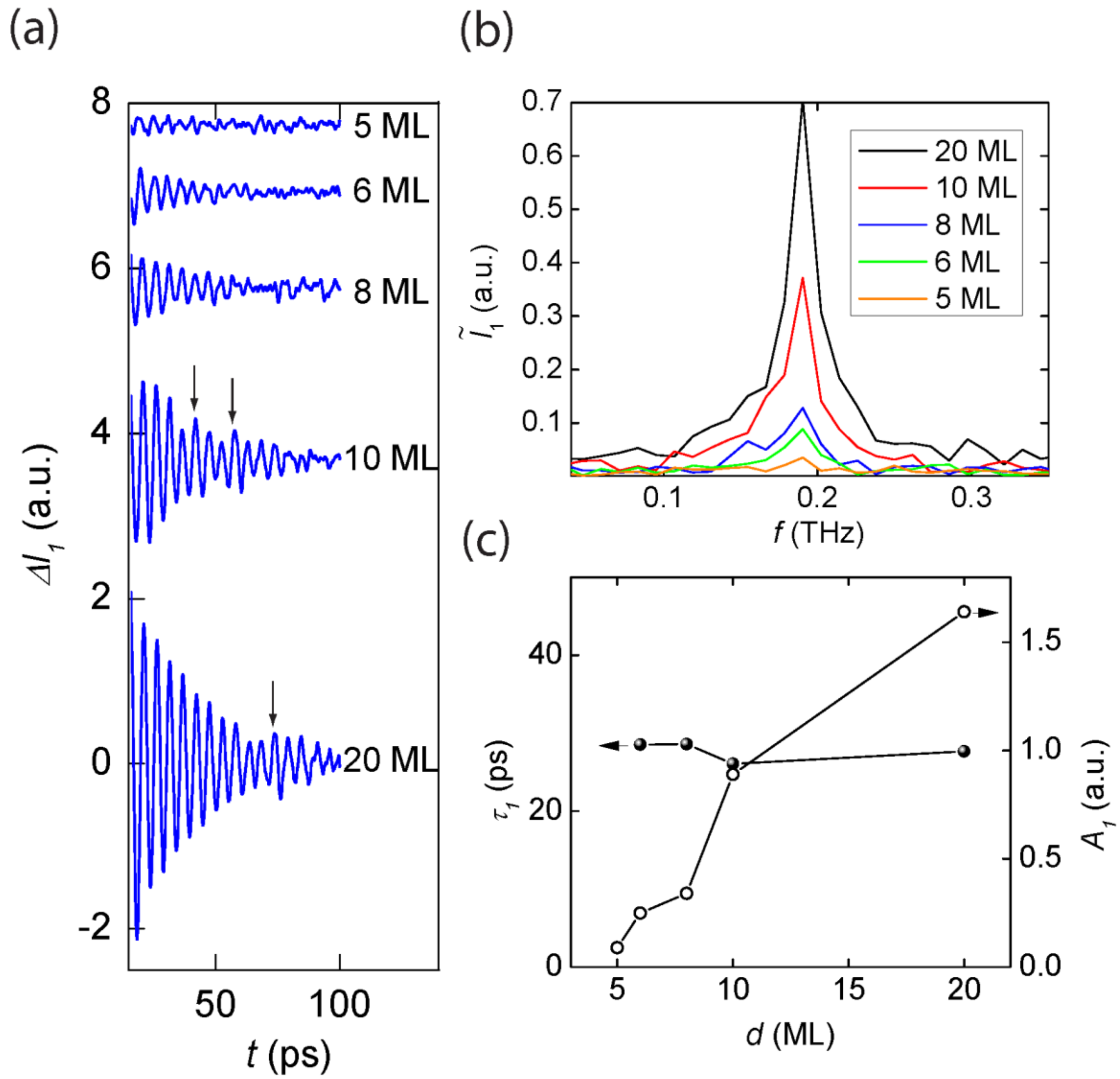

**Figure 3** (a) An under-damped vibrational mode $\Delta I_1$ observed in $SrTiO_3$/Si thin films. Arrows point to places where signals deviate from a simple damping function. (b) Fourier transform (FT) of the oscillating mode indicates a phonon frequency $f_I$ = 190 GHz for all the $SrTiO_3$ film thicknesses measured. (c) As the $SrTiO_3$ film becomes thicker, the measured amplitude of the phonon oscillation $A_I$ increases while the decay constant $\tau_I$ remains approximately the same.

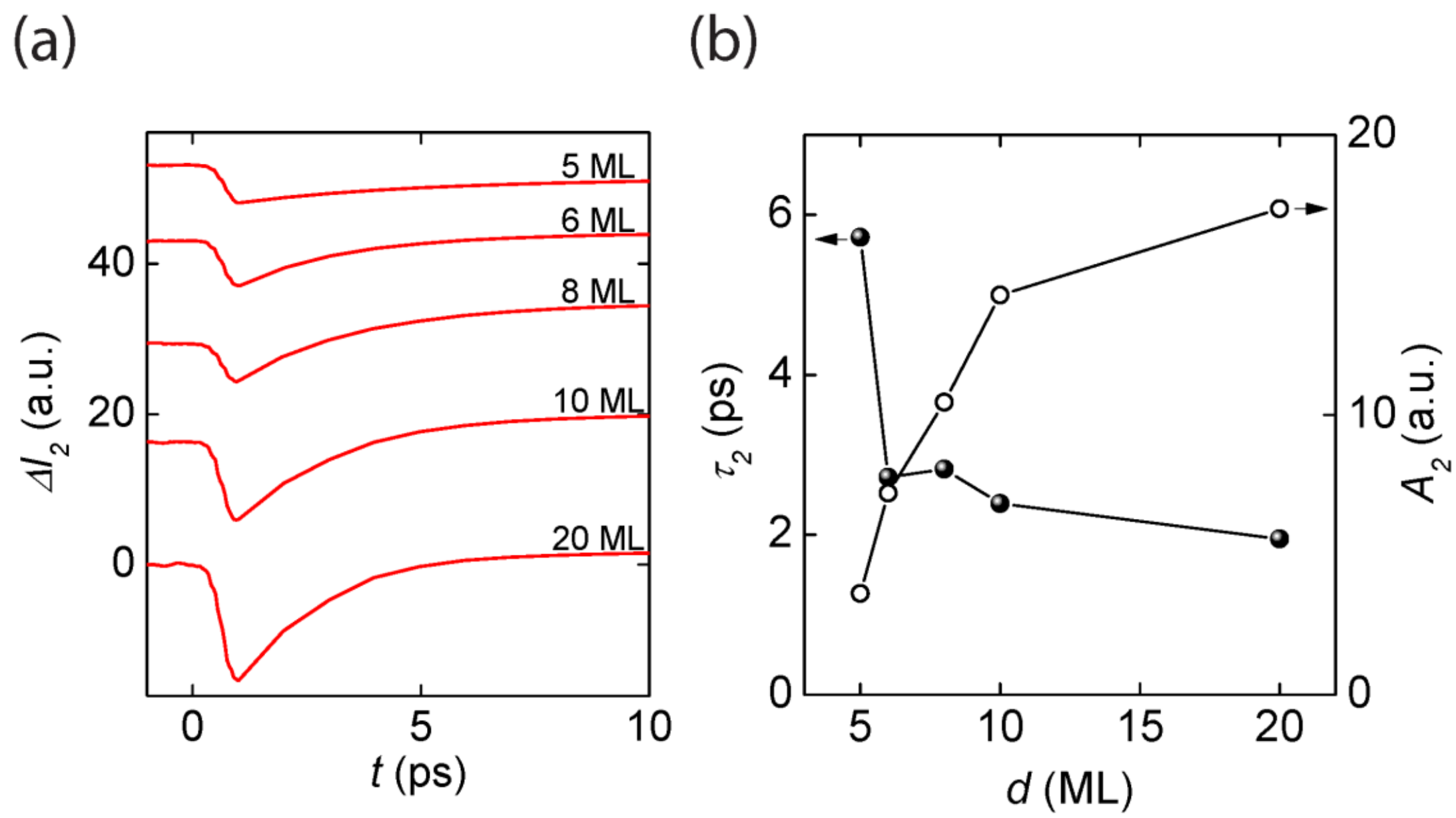

**Figure 4** (a) An over-damped phonon mode $\Delta I_2$ observed in $SrTiO_3$/Si thin films. (b) As the $SrTiO_3$ film becomes thicker, the measured amplitude of the overdamped mode $A_2$ increases while the decay constant $\tau_2$ decreases.